# Processing of massive Rutherford Back-scattering Spectrometry data by artificial neural networks

Renato da S. Guimarães[a], Tiago F. Silva[a1], Cleber L. Rodrigues[a], Manfredo H. Tabacniks[a], Simon Bach[b], Vassily V. Burwitz[b], Paul Hiret[b], Matej Mayer[b]

[a]Physics Institute of the University of São Paulo, São Paulo, Brazil.
[b]Max-Planck-Institut für Plasmaphysik, Garching, Germany.

**Abstract:** Rutherford Backscattering Spectrometry (RBS) is an important technique providing elemental information of the near surface region of samples with high accuracy and robustness. However, this technique lacks throughput by the limited rate of data processing and is hardly routinely applied in research with a massive number of samples (i.e. hundreds or even thousands of samples). The situation is even worse for complex samples. If roughness or porosity is present in those samples the simulation of such structures is computationally demanding. Fortunately, Artificial Neural Networks (ANN) show to be a great ally for massive data processing of ion beam data. In this paper, we report the performance comparison of ANN against human evaluation and an automatic fit routine running on batch mode. 500 spectra of marker layers from the stellarator W7-X were used as study case. The results showed ANN as more accurate than humans and more efficient than automatic fits.

# Introduction

Ion Beam Analysis (IBA) comprises a set of well established analytical techniques for material analysis [1]. Even though some of these techniques are decades old, IBA techniques can still keep their relevance due to many important features hardly found on competing techniques. The IBA technique Particle Induced X-ray Emission (PIXE), for instance, can be performed with an excellent Lower Detection Limit (LDL) when compared to its direct competitor: X-ray Fluorescence (XRF). Rutherford Back-scattering Spectrometry (RBS) provides excellent depth information for thin-film characterization, but its real strength lies in its accuracy, which under certain circumstances, can be close to 1% [2]. On top of that, we can add the fact that RBS has the potential to be a primary standard, i.e. no reference sample is necessary, and absolute quantification is achievable [3].

The point is that these techniques are exceptional and unique. They can provide valuable information on many sorts of samples with essential results for many fields of sciences.

1    Corresponding author: tfsilva@if.usp.br

However, the analysis of measured data is a problem: Besides the fact that it requires a not so simple personnel training, it is also challenging to keep the high standards of quality and traceability when the number of samples and/or measured spectra reaches the scale of hundreds or thousands. For a skilled person, it is evident how to perform the high-quality standard analysis for just a few measurements. However, the protocols to handle a massive number of samples are not entirely clear and raise questions on how to keep consistency along with the whole dataset.

Of course, performing the data processing by a fitting procedure running in batch mode is always a way to go for solving this issue. There are some problems, however, when the analysis requires some level of learning process, which is not possible to solve with typically used fitting algorithms. Additionally, data processing in batch mode always requires some computing time necessary for the fit to converge. This computing time is independent of the number of samples already analyzed. Therefore, the adoption of algorithms capable of learning from experience can, in principle, make the analysis more efficient (especially for similar types of samples) and at the same time, can be a way to ensure the same quality standards for the entire dataset, i.e. the same correctness and bias.

In this study, we compare the performances of human-made evaluations and automatic fits running in batch mode with the performance of an Artificial Neural Network (ANN) when evaluating the same dataset. For this, we took the real problem of the analysis of marker layers deposited on tiles for the internal walls of the W7-X fusion reactor vessel. This dataset consists of 500 spectra from 132 samples analyzed by RBS, which is a number already significantly hard to process. The ANN performance can excel human performance either in processing time or in the quality of the results, and can be a serious candidate for processing large amounts of data, thus improving the RBS throughput.

# Methods

Algorithms are attractive due to their ability to automate and improve performance in tedious and time-consuming tasks. Algorithms can be divided into two categories: the first comprises computer codes directly coded by a human to perform some task; the second has artificial intelligence that delivers a substantial amount of inference derived from observations and training [4].

The popularity of artificial intelligence has increased considerably in the past decade as more and more areas have introduced this concept. Among the many algorithms, ANN has been around for quite some time. It consists of a system composed of many simple processing elements or units, operating in parallel and whose function is determined by the architecture, connection strengths (synaptic weighs), and the processing performed at each node [5]. The massively parallel distributed processor has a natural propensity for acquiring, storing experiential knowledge and making it available for further use [6], with each element only operating on local information and asynchronously [7].

A few works reported the use of ANNs applied to RBS spectra processing [8] [9] [10], showing that it is possible to analyze RBS spectra with neural networks. ANN can thus be useful to substitute the task of analyzing data and tackle in both ways a time-consuming task and assuring consistency for every analysis.

Almost 500 spectra from samples of marker layers deposited on tiles for the W7-X stellarator are used in our case study. These are composed of a 5 - 10 µm thick carbon film on top of a 200 nm thick molybdenum film, all on top of a carbon bulk substrate. This layer structure is designed for an experiment targeting the analysis of erosion due to plasma exposure in the stellarator. By measuring the thickness of the top carbon film before and after the plasma exposure the erosion pattern is revealed [11]. Here, we focus on the analysis procedure of the 500 samples before the plasma exposure as a case study for checking the performances of ANNs against the human evaluation and fitting in batch mode.

## RBS and ANN

RBS is an analytical technique that employs mono-energetic ion beams, typically within a 0.3 – 3 MeV/u energy range, to determine the atomic composition of materials. Since the incident beam energy is well known, measuring the energy ratio of the back-scattered particles ($E_{out}/E_{in}$) gives information about the species of the target nuclei. The previous knowledge of cross-sections (scattering probabilities) enables the determination of the atomic density in the material. Since the ion continuously looses its energy as it penetrates deeper into the target, it is also possible to determine at which depth the collision occurred and thus produce a composition profile in depth of the sample.

The physics involved in RBS can be modeled with excellent precision [12] using first principles and assuming only classic scattering with a central force field [13]. SIMNRA is a software that simulates theoretical RBS spectra given the sample description and setup parameters. Because it is analytical, it can run several orders of magnitude faster than Monte-Carlo based codes, showing excellent agreement for small and medium energy losses [14]. The description of physics models involved in SIMNRA is available in previous works [15] as well as its application to self-consistent analysis while minimizing an objective function [16].

The problematic point is that, even with a relatively simple physics model, the interpretation of an RBS spectrum is not simple: ambiguities and signal overlap often lead to difficulties understanding or even to misinterpretations. Thus, the analysis of a complex RBS spectrum is a challenge to an experienced analyst and also to ANNs. In this study, we target testing the ANN features as a tool to be used to analyze samples with some level of complexity. The large number of samples processed here offers an interesting view on ANN performance compared to other data processing methods.

The high roughness of the layers introduces the complexity here. In the analyzed samples, the roughness stays between 15% and 50%. Even for a trained human, roughness can introduce additional difficulty in interpreting RBS spectra since it affects the shape in different ways and multiple locations. For instance, a top layer's roughness affects its signal's shape and the signals' shape from deeper layers. Thus, roughness strongly correlates the outputs, and the correct prediction of the layer thicknesses by the ANN depends on the correct interpretation of the roughness effects on the spectra.

We used SIMNRA to generate the training and validation sets for an ANN designed to process the data from the W7-X tiles. The dataset consists of many spectra uniformly distributed within specified ranges for each parameter of interest. The aim is that the ANN learns by examples generated with SIMNRA what is the influence on the RBS spectra of the variation of each parameter, and after training, make predictions within acceptable levels of accuracy for each of them. Carbon layer thickness, molybdenum layer thickness, oxygen content of the carbon layer, and its roughness are the principal parameters to be learned by the ANN.

## Input and training data

The experimental RBS spectra are histograms of counts with 1024 channels. The first hundred channels always have the same shape for every spectrum as this part of the spectrum originates from ions scattered in the carbon bulk. This part can thus be discarded because it contains no information on the layer structure. The higher energy part of the spectrum can also be discarded due to the lack of signal (see Fig. 2), yielding roughly 800 channels with information regarding the sample layer structure. It is essential to remove the region with no information content from the spectra to simplify the network, reduce the size of the training set, and optimize the training time.

It has been shown in previous works that a re-binning of RBS spectra within certain limits may be performed without significant loss of information with the advantage of reducing the complexity of the network and the size of the training set [8], [9]. Therefore, we re-binned the spectra by every two channels to decrease the size of the first layer in our neural network, yielding approximately 400 channels sized reduced spectra.

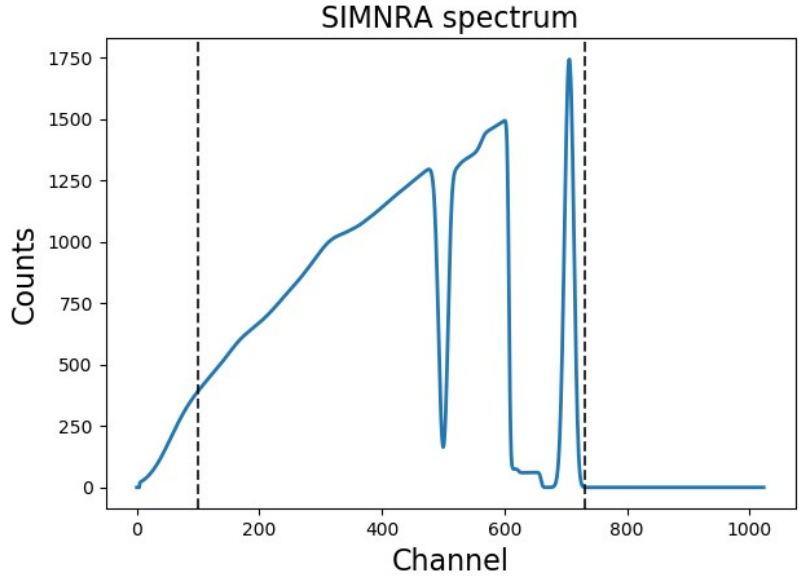

Figure 1 – A typical SIMNRA generated spectra of a carbon layer on top of a molybdenum layer over a substrate of carbon. The region of interest is marked by dashed lines.

After generating the training set with SIMNRA (more details ahead in the text), we added Poisson noise to each spectrum and implemented the same cuts and re-binning as mentioned above. A scaling transformation is also necessary; thus, the average ($X_i$) and standard deviation ($\sigma_i$) for each channel are calculated, and the scaling of the data according to Eq. 1 is performed. One should notice that since the height of the spectra contains information about the deposited charge during the measurements, normalizing all spectra by area could be performed as a form to mitigate its influence in the ANN training. The typical reduced spectra, after scaling and re-binning, may be seen in Fig. 3.

$$x_i^{new} = \frac{x_i^{old} - X_i}{\sigma_i} \quad\quad\quad \text{Eq. 1}$$

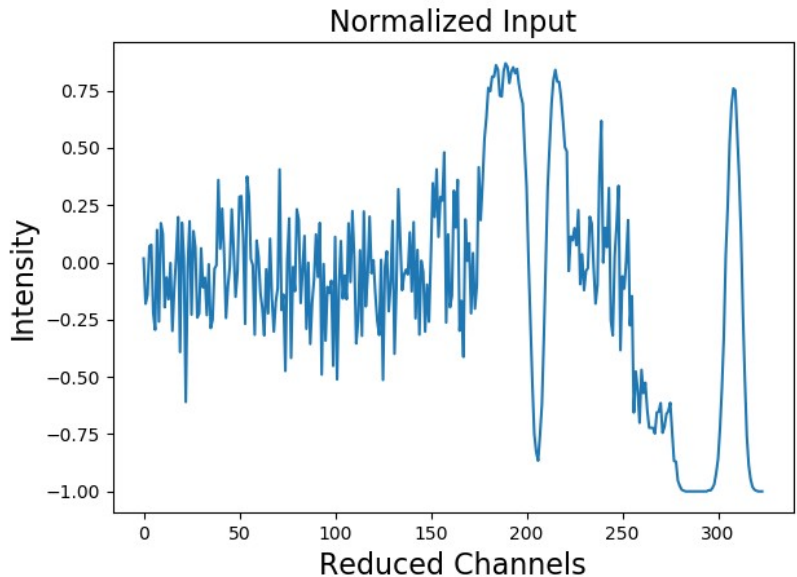

Figure 2 – Reduced and scaled spectra, following the formula presented at Eq. 1 and a dataset of 300,000 samples.

It is important to notice that once the scaling is performed and the mean and standard deviation of the channels are calculated, the same transformation, with the same values, shall be used for scaling the experimental data.

## ANN architecture

It has been shown that any continuous function within an n-dimensional cube may be approximated by the superposition of functions of one variable and by sums of functions [17]. A discussion how this theorem is reproduced by multi-layer perceptrons to approximate any given function to an arbitrary precision has been done in [18]. It was also shown that the lack of success is linked to inadequate learning, insufficient number of hidden layers or lack of deterministic relationship between input and desired output.

Thus, we performed some experiments in an attempt to determine what is the best architecture for the ANN suited to this particular RBS spectra. The quality was defined not just in terms of accurate prediction, but also in terms of how fast an architecture can be trained. For a general purpose ANN there is no constraint for $m \leq n$ (with $m$ being the number of parameters in the output and $n$ the number of input variables), but one should note that the network's ability to represent an m-dimensional vector as output will necessarily be architecture dependent.

As a rule of thumb, every test was performed having the same amount of nodes in the first layer as the number of channels in the spectra, then gradually decreasing it until reaching the desired number of outputs. A proportion of 1, 1/2, and 1/4 times the input array was fixed once it presented the fast learning curve (see Fig. 3) and used for further research. The two hidden layers architecture presented a slower learning curve and the four hidden layer architecture presented some oscillations in the learning curve, probably due to some problem in the implementation of the training algorithm.

The Loss function presented in Fig. 3 represents the deviation between the predictions done by the network and the validation set accordingly to the training step (epoch). This can be interpreted as the lower the loss the higher the learning by the ANN.

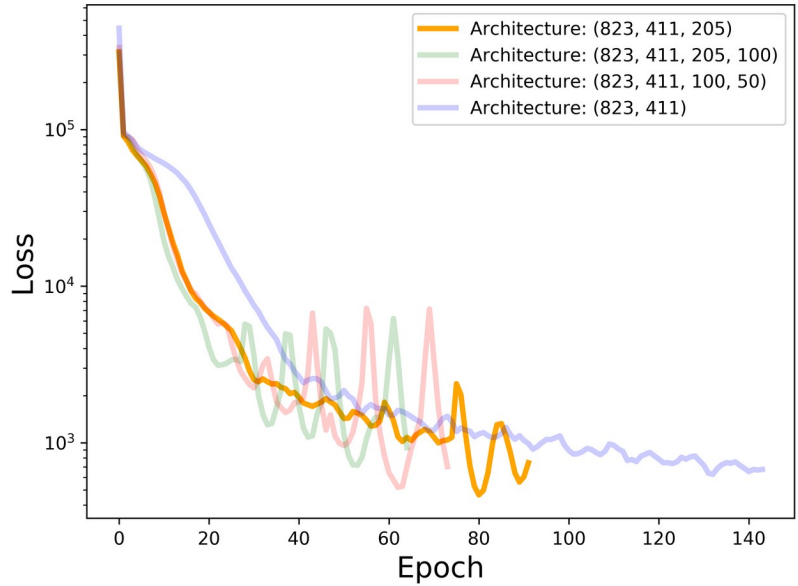

Figure 3 - Loss function (squared-loss) of neural networks trained in different architectures. Each network learned from the same data set, split randomly into 75-25% for training and validation, respectively. The training termination criteria was the loss function not improving by $10^{-4}$ for 10 consecutive iterations. The orange represent the chosen architecture, the legend shows the amount of neurons in each hidden layer, not considering the input or output layers.

## Automation

A Python routine developed by us to control SIMNRA through its OLE functionality is schematically depicted in Fig. 4. Using MPI4py module we parallelized the generation of spectra, with one SIMNRA instance per available computing thread. Simple spectra consisting of a carbon bulk, a molybdenum intermediate layer and a carbon/oxygen surface layer were generated at a rate of roughly 60 spectra/second. Including surface roughness the generation time decreased to about 12 spectra/second.

The samples used for training and validation were generated by SIMNRA uniformly distributed within the specified range for each parameter. About 300,000 spectra were generated for the final run of the ANNs, with and without surface roughness, taking over a week of computer time on a Ryzen 5 2600 processor (six cores and twelve threads). This large amount of samples were generated to study how the training set size influences the final result, Our findings, however, indicate that no significant improvement is obtained with training set sizes larger than 10,000 for this particular case study.

Once the theoretical spectra had noise added, they were reduced and scaled. The ANN training took about 45 minutes with a custom class employing the Scikit-Learn library [19], and to save the ANN state, training and validation set. At the end of the routine, we inserted the ANN prediction back into each SIMNRA file and saved as a prediction file, creating a summary with all said predictions.

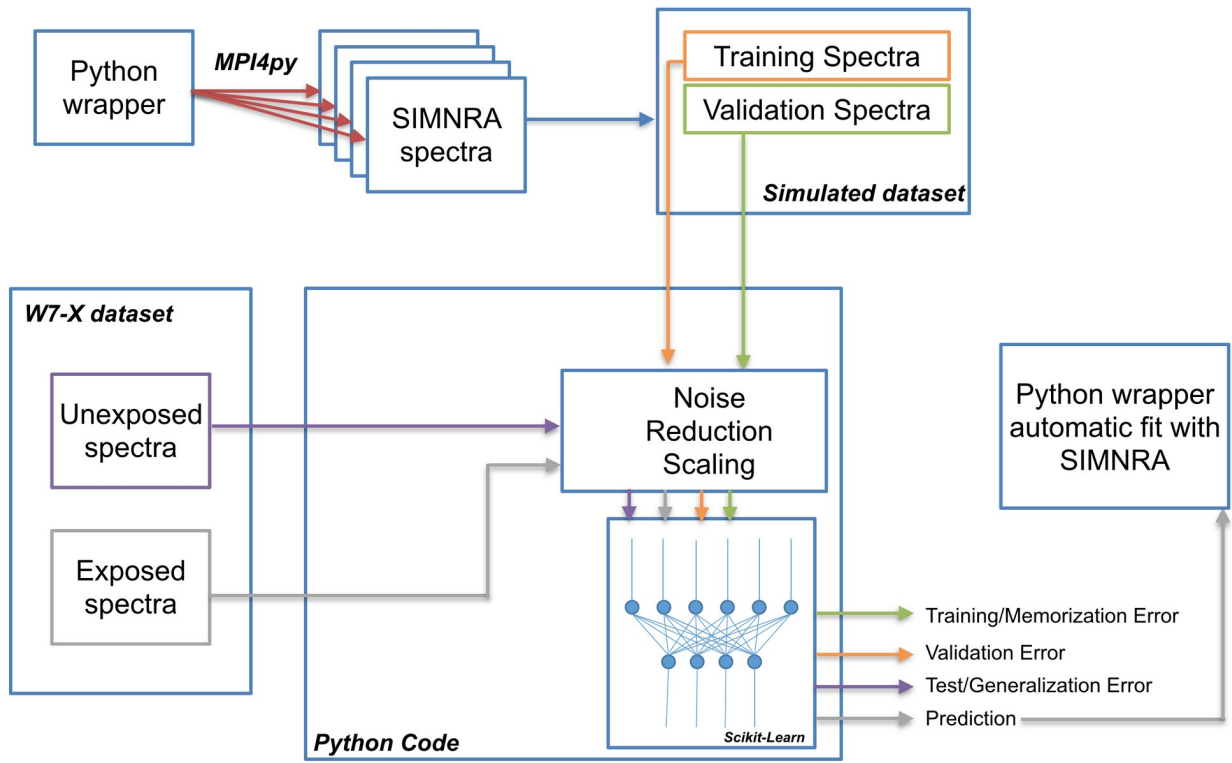

Figure 4 - Schematic for the automated cycle of ANN training and data processing.

# Results and discussion

In order to assess the ANN performances, we compared its predictions to the results of a fit performed by a trained student and by an automatic fit routine. Both methods, the human evaluation and the fit by classical algorithms, are the two widely accepted and promptly approved by any referee. The human evaluation consisted of an eye search of agreement between experimental spectra and simulation. The automatic fit was implemented by successive repetitions of the Nelder-Mead Simplex algorithm as implemented in SIMNRA.

As mentioned, it is not clear what are the protocols to manually keep consistency and the quality standards over an entire dataset with a large number of samples. Our results show that ANN can indeed be consistent. In Fig. 5 we noticed a constant offset between the ANN and the human predictions, even though the network, in this case, was trained with a simplistic model that does not consider roughness in the carbon layer. However, it is possible to observe that this offset is reasonably constant.

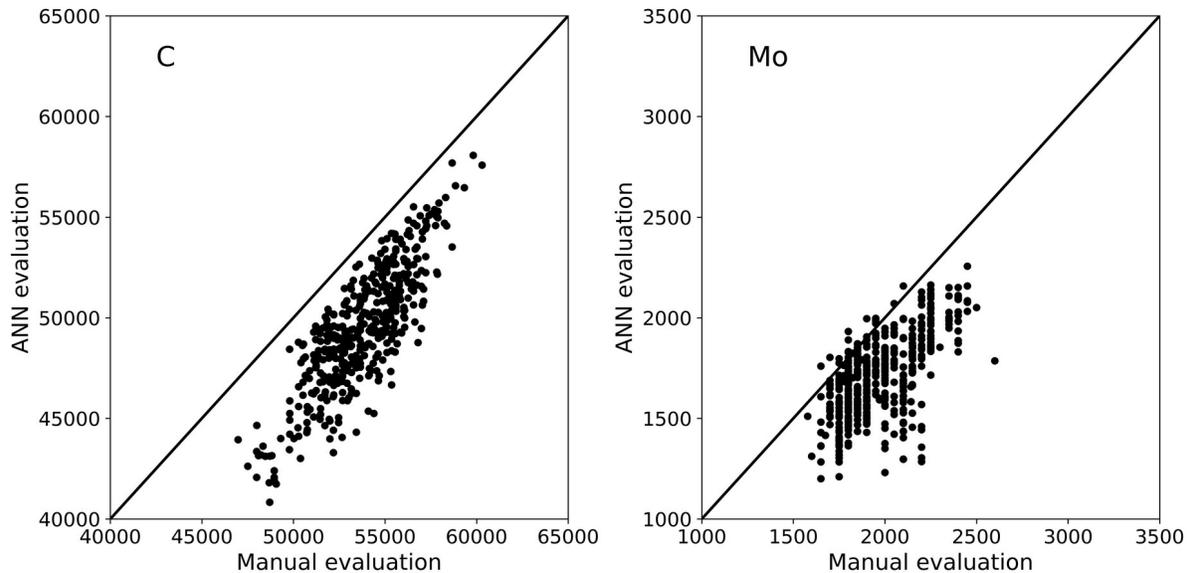

Figure 5 – Carbon and Molybdenum prediction without considering surface roughness of ANN (y-axis) versus Human Evaluation (x-axis). The solid line represents y=x, i.e. the expected outcome assuming no bias from the human agent and full reproduction by ANN. (Training set size: ~300.000).

When the model considers roughness, and the roughness parameter is included as an ANN output, the offset referred to the manual evaluation is reduced considerably. This effect is shown in Fig. 6. Thus, learning the roughness features in the spectra improved the ANN predictions for both carbon and molybdenum.

It is worth to mention that the manual evaluations for the amounts of molybdenum are distributed in vertical lines in both Figs., 5 and 6. This is an evidence of bias in the human evaluation of the data, which is not present in the ANN predictions.

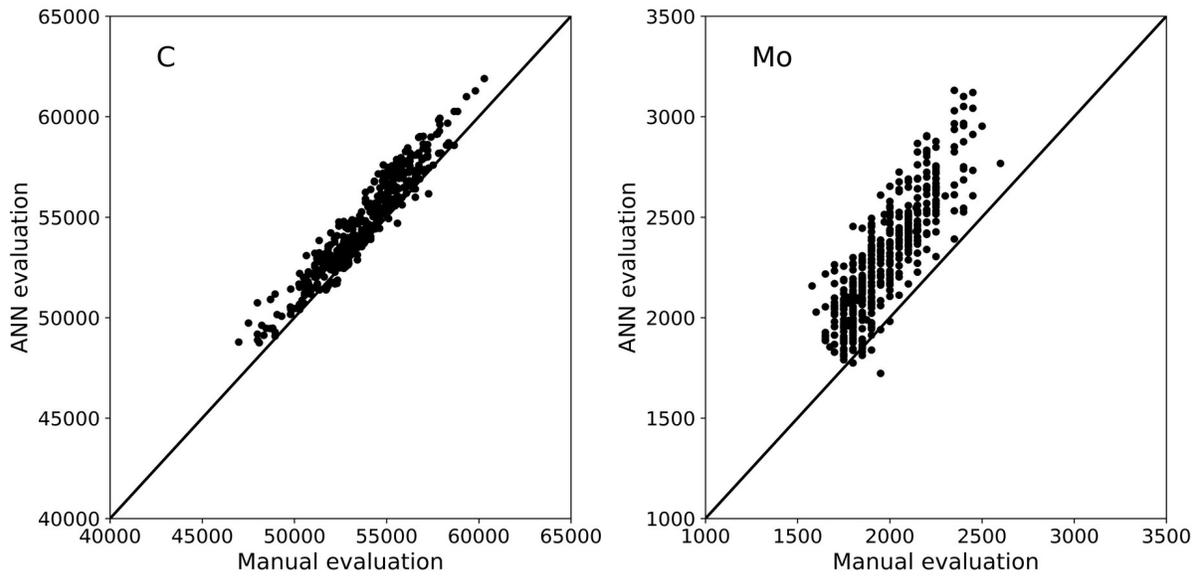

Figure 6 – Carbon and Molybdenum prediction considering surface roughness of ANN (y-axis) versus Human Evaluation (x-axis). The solid line represents y=x, i.e. the expected outcome assuming no bias from the human agent and full reproduction by ANN. (Training set size: ~300.000).

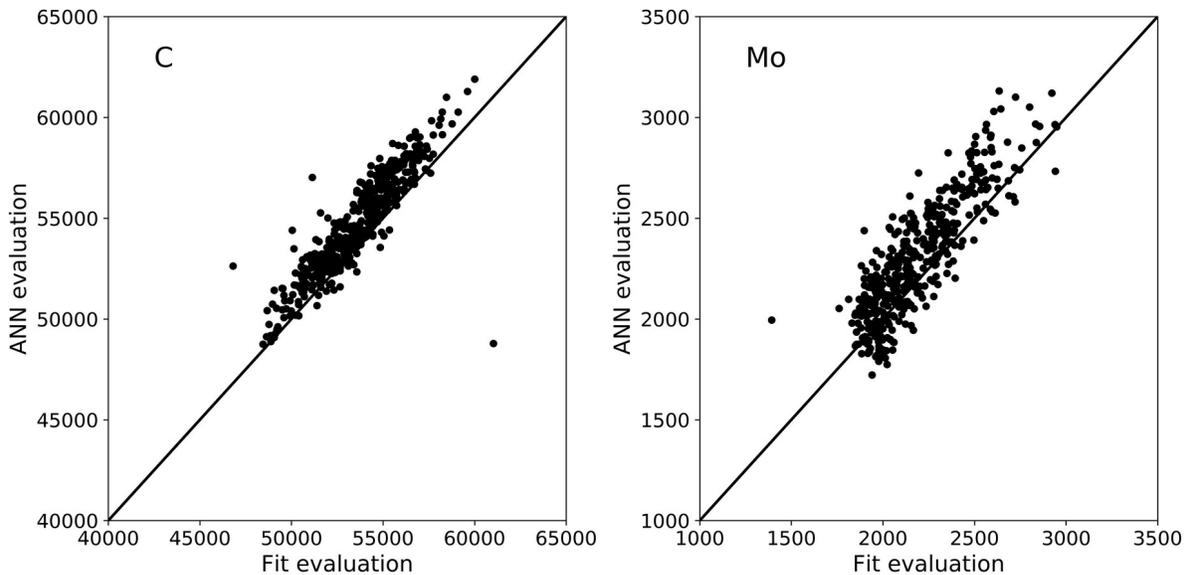

Figure 7 – Carbon and Molybdenum prediction considering surface roughness of ANN (y-axis) versus results of an automatic fit routine evaluation (x-axis) working in batch mode. The solid line represents y=x, i.e. the expected outcome assuming no bias from the automatic fit and full reproduction by ANN. (Training set size: ~300.000).

The comparison of the ANN evaluation with the automatic fit routine is more revealing. The crossing of the cloud of points by the y=x line evidences the excellent agreement of the methods. It is possible to observe a few outliers that are not present in the ANN evaluation nor in the manual evaluation. This is probably because the automatic fit can sometimes get stuck in local minima, while the ANN and the humans have a complex inference ability to recognize and avoid such a situation.

The histogram of the differences between the ANN predictions and the automatic fit evaluations indicate that a small bias is still present: 2% for carbon and 4.8% for molybdenum. The distribution presents a Gaussian-like distribution with a standard deviation of 2% for carbon and 4.8 for molybdenum.

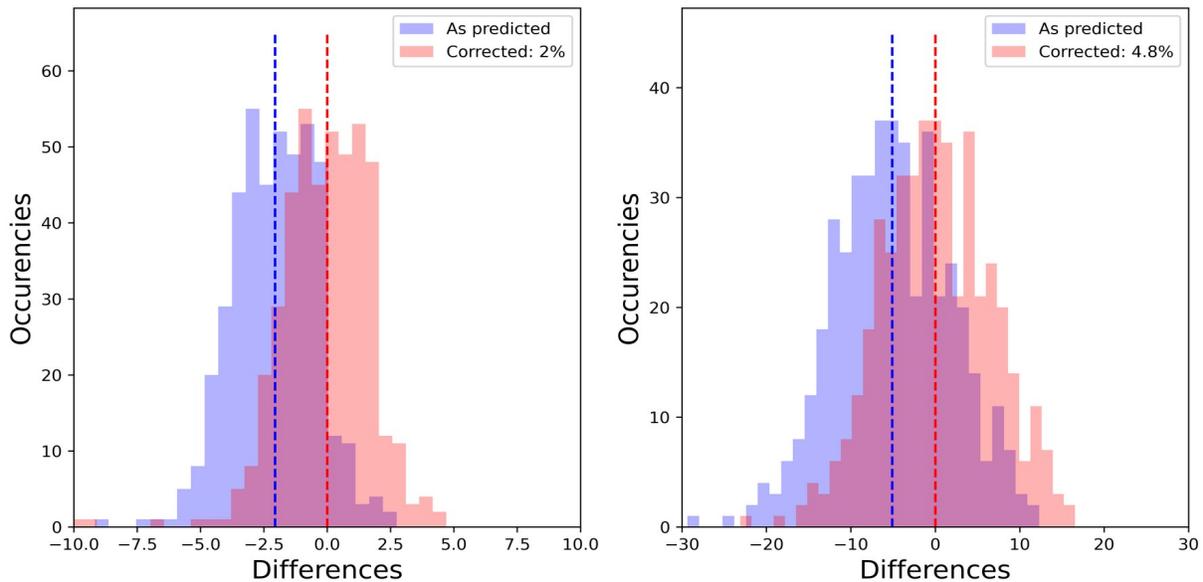

Figure 8 – Distribution of the differences between the ANN predictions and the automatic fit evaluation. A small bias is still present and the corrected distributions are presented.

The differences are positively correlated. We observed a correlation factor 0.37. This indicates that the error in evaluating one leads to the error on the evaluation of the other. One also notices the outliers predicted by the ANN for carbon occur for the same sample as the outliers for molybdenum.

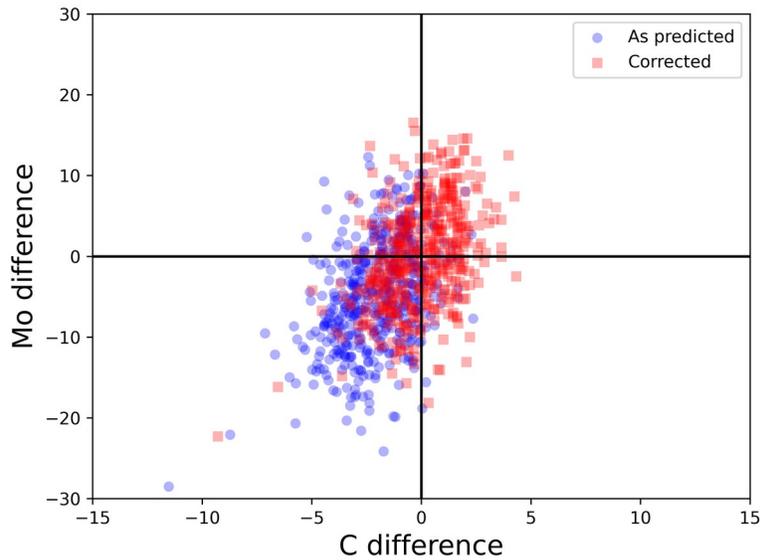

Figure 9 – Scatter plot indicating the correlation of the differences presented in figure 7. A positive correlation is observed.

Since during the training, the ANN initializes weights with random numbers on its connections, we determined the training error for sixty networks trained with the same dataset and architecture. We also checked the influence of the size of the training set by increasing the number of examples used in the various trainings. Fig. 8 presents the result.

Despite some spikes in the root mean squared (rms) of the error for the molybdenum, it stays close to 1%. The rms error for the carbon layer thickness can be even lower. However, the situation is entirely different for the oxygen content: with rms errors reaching 100%, the network is unable to make reasonable predictions for this parameter. The main reason for that is the signal-to-noise ratio for the oxygen signal, which lies between the channels 225 and 275 in Fig. 3. Thus, the noise damages the ANN evaluation of trace elements, i.e. of elements with bad signal-to-noise ratio.

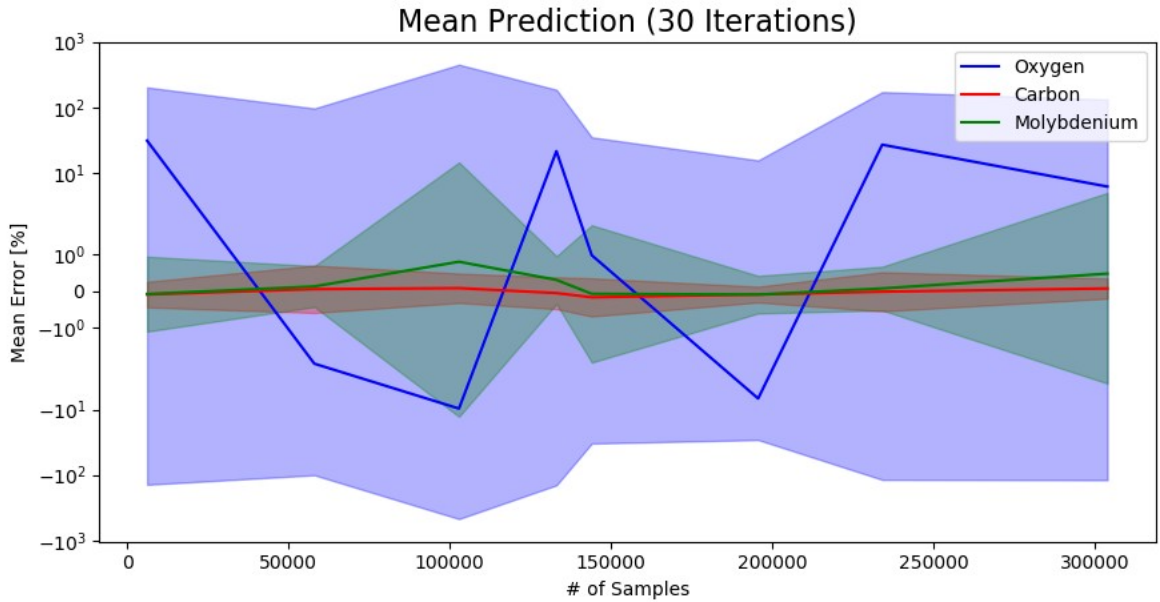

Figure 8 – Mean training error evolution of 60 ANNs trained for different training-set sizes, with learning truncated at 30 interactions. Solid lines represent the mean error while the filled portions represents its root mean squared.

## Conclusions

We evaluated the performance of ANN to process a massive number of RBS spectra. Almost 500 spectra were used as a study case to validate the use of artificial intelligence by comparison with other methods: human evaluation and automatic fit routine running in batch mode. These two methods are widely used and promptly accepted. The ANN faces some preconceptions regarded as "black boxes".

The ANN showed to be more accurate than humans and more efficient than automatic fits with classical algorithms, with comparable results. The better accuracy than humans became evident with Fig. 6. The efficiency issue becomes evident by comparison of the total time spent in the analysis. While the ANN cycle (training set generation, training and data evaluation) took 8 hours of processing, the cycle of automatic fits (convergence of the optimization algorithm) took 4 days.

The evaluation of the error with repetitive training and with different sizes of training sets is important for a complete evaluation of the uncertainties associated to the use of artificial intelligence. The ANN uncertainty could be evaluated as approximately 1% for this case, and if this method of data processing is used, this should be considered in the uncertainty budget.

# Acknowledgments

R.S. Guimarães and T.F. Silva acknowledge the financial supported of University of São Paulo. The authors acknowledge Udo von Toussaint for comments and suggestions.